\documentclass[12pt]{iopart}
\usepackage{graphicx}
\begin{document}
\title{Inhomogeneous spacetimes as a dark energy model}
\author{David Garfinkle}
\address{Dept. of Physics, Oakland University,
Rochester, MI 48309, USA}
\ead{garfinkl@oakland.edu}
\begin{abstract}
Lema\^itre-Tolman-Bondi inhomogeneous spacetimes
are used as a cosmological model for type Ia 
supernova data.  It is found that with certain parameter choices 
the model fits the data as well as the standard $\Lambda$CDM 
cosmology does. 

\end{abstract}
\maketitle
\section{Introduction}

Observations of type Ia supernovae\cite{perlmutter,riess}
have led to the conclusion that 
the expansion of the universe is accelerating.  Since any matter
with non-negative pressure causes a deceleration, 
this leads to the conclusion that the universe contains a substantial
amount of an exotic type of matter refered to as ``dark energy.''  
A cosmological constant can account for the requisite dark energy, 
and indeed a model with a cosmological constant as well as ordinary 
baryonic matter and cold dark matter, the $\Lambda$CDM model, has
become the standard cosmology.  Though it agrees quite well with the
data, the $\Lambda$CDM model has certain disturbing features:  
From the point of view of 
Planck units, the cosmological 
constant seems unnaturally small by about 120 orders
of magnitude.  In addition, the $\Lambda$CDM model requires that
we live in the very special cosmological era when matter and dark
energy have comparable densities.  These disturbing features motivate
the search for alternatives to the standard $\Lambda$CDM model.  One
such alternative is the theory of Kolb {\it et al}\cite{kolb,notari,kolb2} 
that the universe contains only ordinary matter and that the apparent 
acceleration is an effect of the universe's inhomogeneities.  However, 
the work of \cite{kolb,notari,kolb2} has been criticized on many 
grounds.\cite{bob,eanna0,fry}
In particular, their treatment uses perturbation theory beyond
its regime of validity since the effects that they try to account
for are large.  Since a full non-perturbative treatment of 
an inhomogeneous, anisotropic universe is quite difficult, 
some authors\cite{nambu,moffat,iguchi,tetradis,mansouri1,mansouri2}
have treated a toy problem: an inhomogeneous but spherically symmetric 
universe of pressureless matter.  These models, the 
Lema\^itre-Tolman-Bondi (LTB) 
spacetimes,\cite{lemaitre,tolman,bondi}
are well studied, so all that is needed is to apply them to the question of
the apparent acceleration of the universe.  This has been done by 
several authors;\cite{nambu,moffat,iguchi,tetradis,mansouri1,mansouri2} 
however, for the most 
part the authors of 
these references have used purely formal notions of the acceleration
of the universe.  In contrast, Vanderveld {\it et al}\cite{eanna} 
and Alnes {\it et al}\cite{alnes} have emphasized 
that it is more appropriate to see whether the model fits the relation
between redshift and luminosity from which the cosmological acceleration
was deduced.

In this paper, we will treat a class of 
LTB spacetimes and
calculate their compatiblity with the observational data 
of\cite{perlmutter}.  Section 2 contains a presentation of the notation
and methods to be used.  Results are presented in section 3 and their
implications discussed in section 4.

\section{Luminosity distance in LTB models}      

The LTB metric takes the form\cite{landau}
\begin{equation}
d {s^2} = - d {t^2} + {\frac {{r'}^2} {1 + f}} d {{\tilde r}^2}
+ {r^2} (d {\theta ^2} + {\sin ^2} \theta d {\phi ^2} ) 
\end{equation}
Here the area radius $r$ is a function of both coordinates $t$ and
$\tilde r$ and a prime denotes derivative with respect to $\tilde r$.  
The function $f$ depends only on $\tilde r$.   It follows
from Einstein's equation that 
\begin{equation}
{\dot r}^2  = f + {\frac F r}  
\label{rdot}
\end{equation}  
where an overdot denotes derivative with respect to $t$ and where the function
$F$ depends only on $\tilde r$.  
The density is given by
\begin{equation}
\rho = {\frac {F'} {8\pi {r^2} {r'}}}
\end{equation}

It is helpful to introduce the quantities $a, \, A$ and $B$ by
\begin{eqnarray}
r &=& a {\tilde r} \\
f &=& A {\tilde r}^2 \\
F &=& B {\tilde r}^3
\end{eqnarray}
Then equation (\ref{rdot}) becomes 
\begin{equation}
{\dot a}^2 = A + {\frac B a}
\label{adot}
\end{equation}
The solution of equation (\ref{adot}) is 
\begin{equation}
t - {t_0} = {\int _0 ^a} {\frac {d u} {\sqrt{A + {\frac B u}}}}
\label{asoln}
\end{equation}
where $t_0$ is a function of $\tilde r$ whose physical meaning is the time 
at which the shell of dust labeled by $\tilde r$ has zero radius.  

At first sight, it seems that the LTB solutions depend on 
three arbitrary functions: $A, \; B$ and $t_0$.  However, there is still
the coordinate freedom to change $\tilde r$ to any function of $\tilde r$.
We will use that coordinate freedom to make $B$ constant.  This makes
the solutions with constant $A$ and zero $t_0$ the standard 
Friedman-Robertson-Walker (FRW) homogeneous and isotropic cosmologies.
We will further choose the value of $B$ to be 4/9 so that the $A=0$
FRW cosmology takes the standard form $a= t^{2/3} $.  In addition we
will consider only models with ${t_0} = 0$.  These models have a genuine
Big Bang singularity in the sense that all dust shells are at zero
radius at the same time. 

Define the null vector $\ell^a$ by
\begin{equation}
{\ell^a} = - {{\left ( {\frac \partial {\partial t}} \right ) }^a}
+ {\frac {\sqrt {1+f}} {r'}} {{\left ( {\frac \partial {\partial 
{\tilde r}}}\right ) }^a} 
\end{equation}
Then a future directed null geodesic takes the form 
${k^a} = - \omega {\ell ^a}$ and the redshift $z$ is given by
$1+z=\omega/{\omega_0}$ where a zero subscript denotes the value of
the quantity at ${\tilde r}=0$.  The luminosity distance is given by
\cite{eanna}
${d_L} = r{{(1+z)}^2} $.

We would like to find the luminosity distance as a function of redshift.
To that end, it is helpful to find how each of these quantities varies
as one goes along the past light cone of the observer.
It follows from the geodesic equation 
that
\begin{equation}
{\ell^a}{\nabla_a} z = {\frac {{\dot r}'} {r'}} (1 + z) 
\end{equation}
Since $d_L$ is a function of $z$ and $r$ we need to know also how
$r$ varies along the past light cone.  We have
\begin{equation}
{\ell^a}{\nabla_a} r = {\sqrt {1+f}} - {\dot r} 
\end{equation}
It is also helpful to know how $a$ varies along the past light cone.  
We have
\begin{equation}
{\ell ^a} {\nabla _a} a = {\sqrt {1+f}} {\frac {a'} {r'}} - {\dot a} 
\end{equation}
Thus along the past light cone we find
\begin{eqnarray}
{\frac {dr} {dz}} &=& {\frac {r'} {(1+z){{\dot r}'}}} (
{\sqrt{1+f}}-{\dot r})   
\label{dzr}  \\
{\frac {da} {dz}} &=& {\frac 1 {(1+z){{\dot r}'}}} (
{\sqrt {1+f}} {a'} - {\dot a}{r'})  
\label{dza}
\end{eqnarray}
 
If we can find all the quantities on the right hand sides of equations
(\ref{dzr}) and (\ref{dza}) as functions of $z, r$ and $a$ then we will
be able to numerically integrate these equations and thus find $d_L$
as a function of $z$.  Since ${\tilde r}=r/a$ we can find any function of
$\tilde r$ and thus given a choice for the function $A$ we can find $f$.
Similarly, we can find $\dot r$ using equation (\ref{rdot}) and  
$\dot a$ using equation (\ref{adot}).  Given ${a'}$ and ${{\dot a}'}$  
we have ${r'}=a+{\tilde r}{a'}$ and 
${{\dot r}'}={\dot a}+{\tilde r} {{\dot a}'}$ so all that remains
then is to find ${a'}$ and ${{\dot a}'}$.  Differentiating
equation (\ref{asoln}) with respect to $\tilde r$ and solving for $a'$
we obtain
\begin{equation}
{a'} = {\sqrt {A + {\frac B a}}} I {A '} 
\label{aprime}
\end{equation} 
Where $I$ is given by
\begin{equation}
I = {\frac 1 2} {\int _0 ^a} {{\left ( A + {\frac B u} \right ) }^{-3/2}}
du
\label{Idef}
\end{equation}
This integral can be done in closed form, and for $A>0$ takes on the value
\begin{equation}
I = {\frac B {2 {A^{5/2}}}} \left [ s \left ( 3 + {\frac {Aa} B} \right )
 - {\frac 3 2} \ln \left ( 
{\frac {1+s} {1-s}} \right ) \right ]
\end{equation}
where 
\begin{equation}
s \equiv {{\left ( 1 + {\frac B {Aa}}\right ) }^{-1/2}}
\end{equation}
Now differentiating equation (\ref{aprime}) with respect to $t$ and
using equations (\ref{adot}) and (\ref{Idef}) we obtain
\begin{equation}
{{\dot a}'} = {\frac {A'} 2} \left [ {{\left ( A + {\frac B a} \right )
}^{-1/2}} - {\frac B {a^2}} I \right ]
\end{equation}

\section{Results}

In \cite{eanna} models with $A=0$ and ${t_0} \ne 0$ were considered,
while \cite{alnes} treat models with both $A$ and $t_0$ nonvanishing.  In
contrast, we will consider models with ${t_0}=0$ and $A \ne 0$.  There
remains the choice of the function $A$.  Observations of the Cosmic Microwave
Background suggest that $\Omega =1$ whereas observations of galaxy clusters
suggest that $\Omega$ in matter is approximately 30\%.  In the context
of the LTB models, this suggest that $A$ should approach zero
at large distances and be positive at smaller distances.  We will 
choose $A$ to have the form
\begin{equation}
A = {\frac 1 {1 + {{(c{\tilde r})}^2}}} 
\end{equation}
where the constant $c$ will be chosen for the best fit with the data.  
Let $\Omega _M$ be the value of $\Omega$ in matter at 
our present time and position.
Then in the LTB models with our choice of $A$ and $B$ we have
\begin{equation}
{\Omega _M} = {{\left ( 1 + {\frac {9a} 4}\right ) }^{-1}} 
\end{equation} 
Thus, given a choice for the value of $\Omega _M$ we find that the current
value of $a$ is given by
\begin{equation}
a = {\frac 4 9} ( {\Omega _M ^{-1}} - 1)
\label{azero}
\end{equation}
The initial values of $r$ and $z$ are zero.
We choose the value of $\Omega _M$ and then use equation (\ref{azero})
to get the initial value of $a$.  We then numerically integrate equations
(\ref{dzr}) and (\ref{dza}) outwards to find the luminosity distance as a
function of $z$.  

In \cite{perlmutter} what is computed is the 
effective magnitude, which is
\begin{equation}
{m_B ^{\rm effective}} = {{\cal M}_B} + 5 \log({H_0} {d_L})
\end{equation}
Here $H_0$ is the Hubble constant and ${\cal M}_B$ is a constant 
to be found by the best fit to an FRW model. In our model, $H_0$ is simply
the initial value of ${\dot a}/a$. 
In \cite{perlmutter} the models are all FRW and the free parameters are 
$\Omega _\Lambda$ and $\Omega _M$.  The best fit, the one with the smallest
$\chi ^2$ is for the model with ${\Omega_\Lambda} = 1.32$ and 
${\Omega_M} = 0.73$.  For any other model, whether FRW or not, one can then 
use the supernova data of \cite{perlmutter} to compute its $\chi ^2$ 
and then subtract off the $\chi ^2$ of that best FRW fit to obtain 
$\Delta {\chi ^2}$ which is a measure of the goodness of fit.  In particular,
the standard $\Lambda$CDM model has ${\Omega _\Lambda} = 0.74$ and 
${\Omega _M} = 0.26 $ and has $\Delta {\chi ^2} = 1.2$.

We will consider two different models: one with ${\Omega_M} = 0.3 $ 
and one with ${\Omega _M} = 0.2$.  
In each case the value of $c$ will be chosen 
for best fit with the supernova data.  The errors in both 
$m_B ^{\rm effective}$ and $z$
are used to compute $\chi ^2$.   
\begin{figure}
\includegraphics{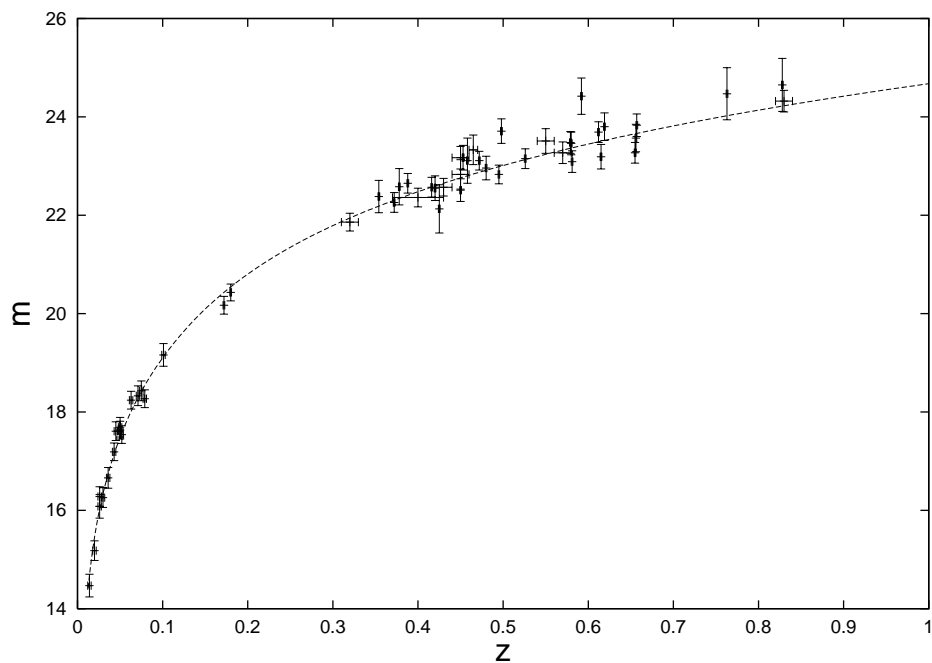}
\caption{\label{fig1}Plot of effective magnitude {\it vs} redshift for the
${\Omega _M}=0.3$ LTB model (curve) and the supernova data.}
\end{figure}
In figure (\ref{fig1}) are plotted the supernova data and the 
${\Omega _M} =0.3$ LTB model.  For this model $c=8.5$ and  
$\Delta {\chi ^2} = 3.8$.  Thus, the standard $\Lambda$CDM model 
(the one with ${\Omega _\Lambda} =0.74$ and ${\Omega _M}=0.26$) is a 
better fit to the data.  However, given the possiblity of systematic
uncertainties in the supernova data, it not clear whether this difference
is significant.  In figure (\ref{fig2}) the ${\Omega_M}=0.3$ LTB
model and the standard $\Lambda$CDM model are plotted.  Note that the
curves agree very closely.  
\begin{figure}
\includegraphics{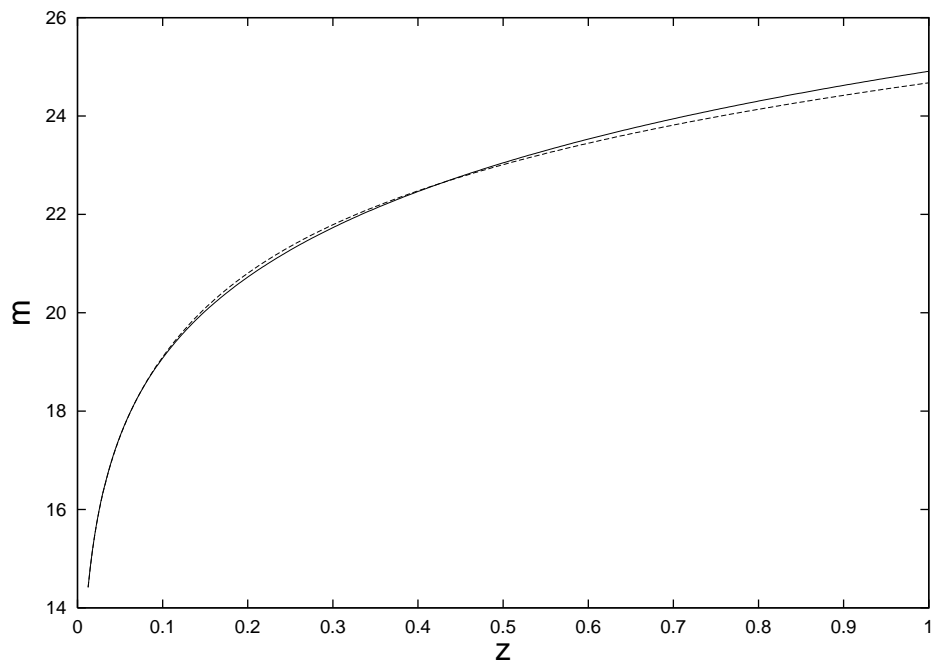}
\caption{\label{fig2}Plot of effective magnitude {\it vs} redshift for the
standard $\Lambda$CDM model (solid curve) and the
${\Omega _M}=0.3$ LTB model (dashed curve).}
\end{figure}
 
\begin{figure}
\includegraphics{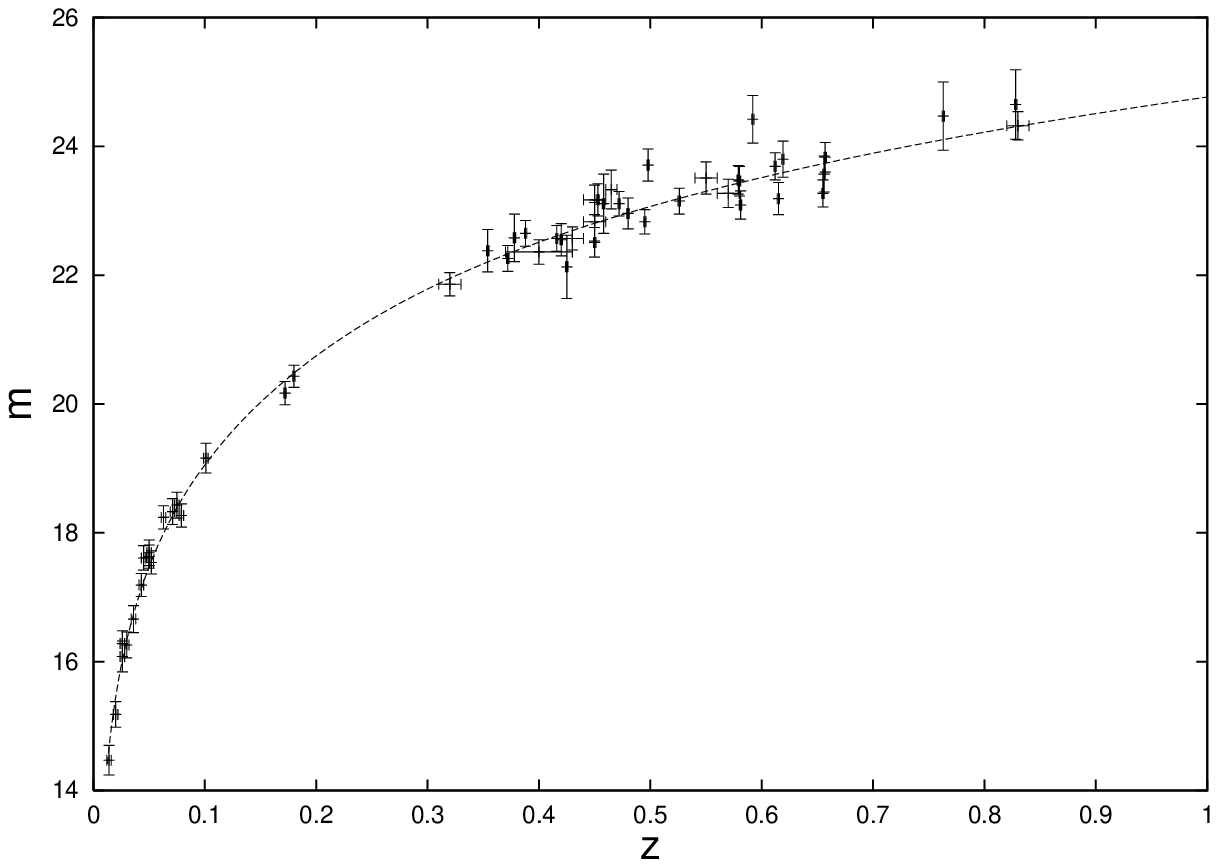}
\caption{\label{fig3}Plot of effective magnitude {\it vs} redshift for the
${\Omega _M}=0.2$ LTB model (curve) and the supernova data.}
\end{figure}

\begin{figure}
\includegraphics{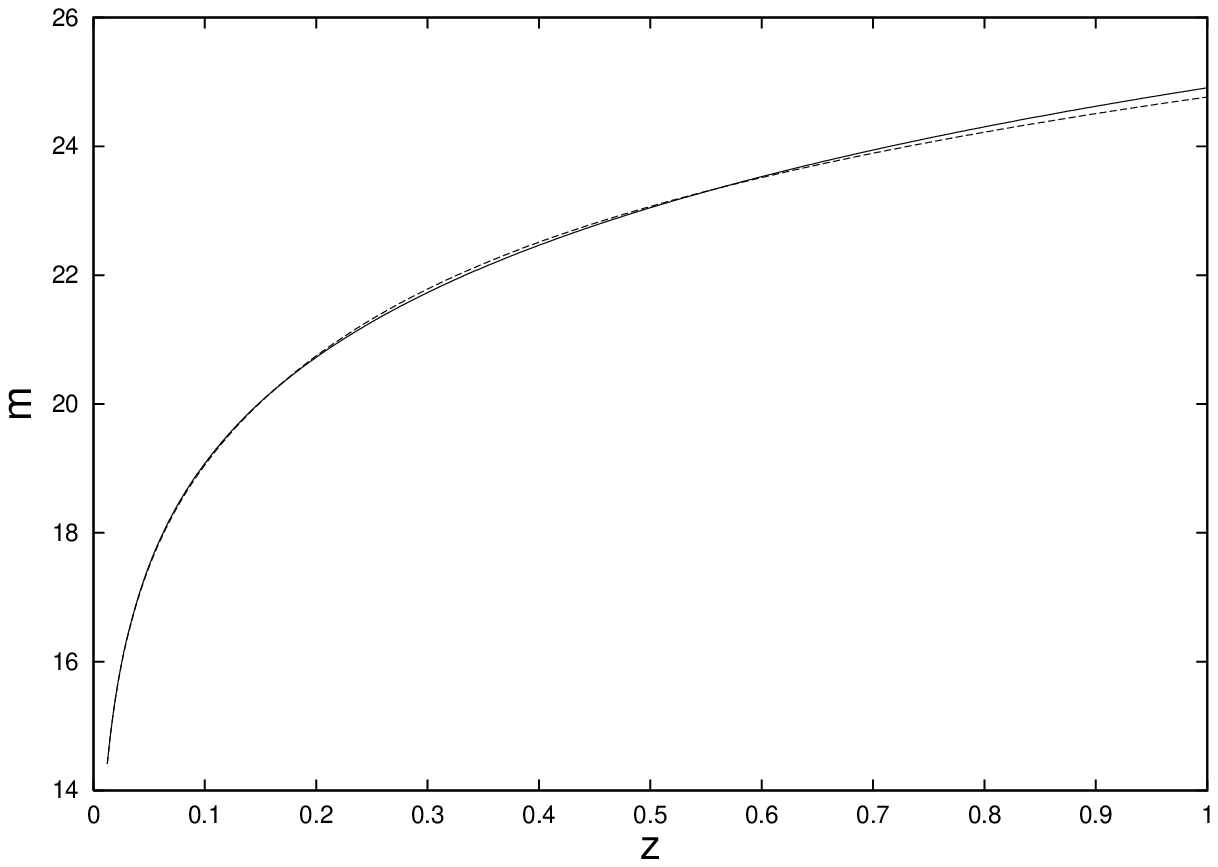}
\caption{\label{fig4}Plot of effective magnitude {\it vs} redshift for the
standard $\Lambda$CDM model (solid curve) and the
${\Omega _M} =0.2$ LTB model (dashed curve).}
\end{figure}

In figure (\ref{fig3}) are plotted the supernova data and the
${\Omega _M} =0.2$ LTB model.  For this model $c=5.1$ and  
$\Delta {\chi ^2} = 0.3$.  This is an even better fit to the data than the 
standard $\Lambda$CDM model.
In figure (\ref{fig4}) the ${\Omega_M}=0.2$ LTB
model and the standard $\Lambda$CDM model are plotted.  Note that the
curves are almost identical.

\begin{figure}
\includegraphics{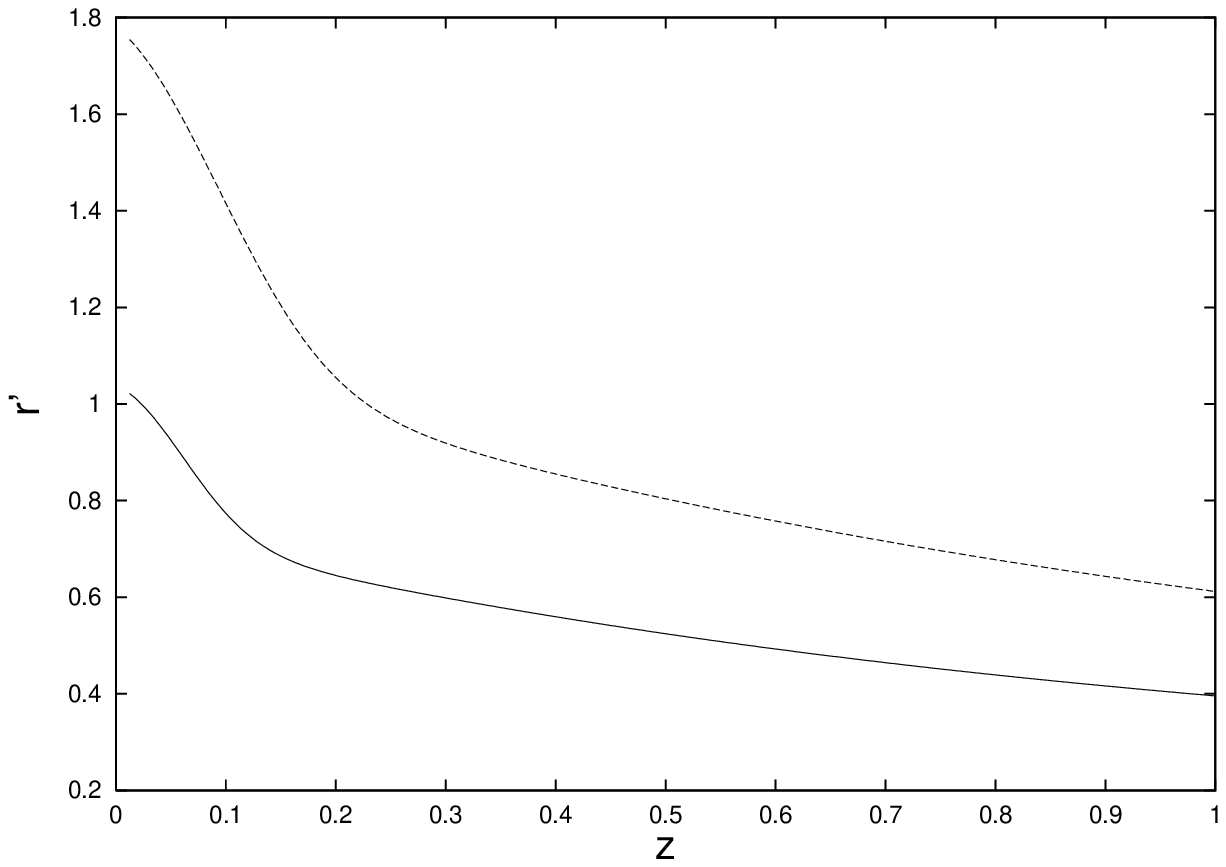}
\caption{\label{fig5}Plot of $r'$ {\it vs} redshift for the
LTB models with ${\Omega _M}=0.3$ (solid curve) and the
${\Omega _M} =0.2$ LTB model (dashed curve).}
\end{figure}

\begin{figure}
\includegraphics{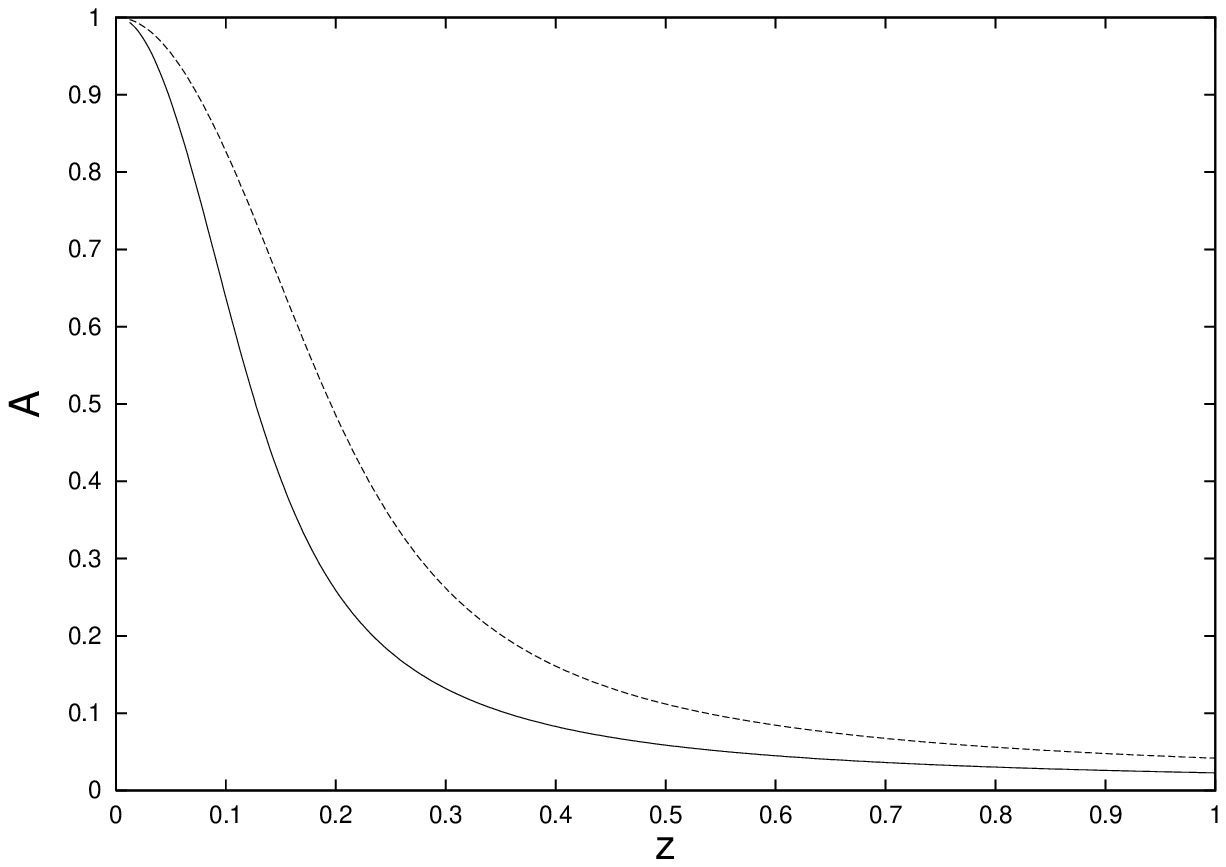}
\caption{\label{fig6}Plot of $A$ {\it vs} redshift for the
LTB models with ${\Omega _M}=0.3$ (solid curve) and the
${\Omega _M} =0.2$ LTB model (dashed curve).}
\end{figure}

In \cite{eanna} it is emphasized that a physically reasonable 
LTB model must be smooth both at the origin and elsewhere.
For the coordinates that we have chosen, smoothness at the origin is
insured provided that $A$ is a smooth function of ${\tilde r}^2$.  
Smoothness elsewhere requires that there be no "shell-crossing" 
{\it ie} that $r'$ not vanish.  In figure (\ref{fig5}) are plotted
$r'$ {\it vs} redshift along the past light cone for the ${\Omega _M}=0.3$ 
and ${\Omega _M}=0.2$ LTB
models.  Since $r'$ does not vanish, there is no shell crossing.  

The fairly large values of the constant $c$ in these models means that
the spatial size of the underdensity in which we live is fairly small.  In
figure (\ref{fig6}) are plotted
$A$ {\it vs} redshift along the past light cone for the ${\Omega _M}=0.3$
and ${\Omega _M}=0.2$ LTB
models.  Note that $A=0$ corresponds to an $\Omega =1$ FRW spacetime.  Thus
in these models, we inhabit an underdensity of fairly small spatial extent in 
what is otherwise a spatially flat FRW cosmology.

\section{Conclusions}

This work shows that an inhomogeneous but spherically symmetric 
cosmological model can account for the apparent acceleration in 
the supernova data without any dark energy.  The models have a
simple form for the inhomogeneity and they fit the supernova
data as well as the standard FRW model with a cosmological 
constant does. 

However, this does not make the LTB spacetime a viable
cosmological model.  There is more cosmological data than the
supernovae.  In particular, data from the cosmic microwave background
and from weak lensing indicate that the universe contains a large amount
of matter that is "unclumped."  This would be accounted for by a 
cosmological constant; but is unlikely to be accomodated by a LTB 
model or any simple modification of one.  Furthermore, since in these 
simple models, we inhabit an underdensity of farily small spatial extent,
the models may not be compatible with data for galaxy counts as a function
of redshift. 
Nonetheless, the work of\cite{alnes} suggests that it might be possible
to accomodate some of this additional cosmological data within the
LTB models.
These issues needs further study.

\ack

I would like to thank \'Eanna Flanagan and Bob Wald for helpful discussions.
This work was supported by NSF grant PHY-0456655 
to Oakland University.

\end{document}